\documentclass[review]{elsarticle}

\usepackage{hyperref}

\usepackage{amsmath}
\usepackage{amssymb}
\usepackage{bm}
\usepackage{multirow}
\usepackage{array}

\usepackage{textcomp}

\usepackage{color}

\journal{Journal of the Mechanics and Physics of Solids}

\begin{document}
\begin{frontmatter}

\title{Simulation of macroscopic systems with non-vanishing elastic dipole components}

\author[CEADEN]{T.~Jourdan}
\ead{thomas.jourdan@cea.fr}

\address[CEADEN]{DEN-Service de Recherches de M\'etallurgie Physique, CEA, Universit\'e Paris-Saclay, F-91191, Gif-sur-Yvette, France}

\begin{abstract}
To simulate a macroscopic system from a simulation cell, a direct summation of the elastic fields produced by periodic images can be used. If the cell contains a non-zero elastic dipole component, the sum is known to be conditionally convergent. In analogy with  systems containing electric or magnetic dipoles, we show that the sum introduces a component which only depends on the shape of the summation domain and on the dipole density. A correction to the direct summation is proposed for the strain and stress fields in the simulation cell, which ensures that zero tractions are imposed on the boundary of the macroscopic system. The elastic fields then do not depend anymore on the shape of the domain. The effect of this correction is emphasized on the kinetics of dislocation loop growth by absorption of point defects. It is shown that correcting elastic fields has an influence on the kinetics if defects have different properties at stable and saddle points.
\end{abstract}


\end{frontmatter}

\section{Introduction}
\label{sec:introduction}

Elastic fields produced by microstructural defects such as dislocations, grain boundaries and precipitates are known to deeply affect the evolution of material properties. Simulations including such elastic effects can be dislocation dynamics~\cite{Amodeo1990,Kubin1992,Ghoniem1999a,Arsenlis2007}, phase field~\cite{Rodney2003,Chen2002} and object kinetic Monte Carlo~\cite{Subramanian2013,Vattre2016}. Whatever the method used, precise values of the elastic field produced by the microstructural defects must be determined. To simulate a large system, it is common to use periodic boundary conditions (PBCs) and the influence of defects outside the simulation box must be taken into account for the calculation of the elastic fields. If elastic fields are calculated by solving the mechanical equilibrium in Fourier space, this contribution is taken into account naturally~\cite{Rodney2003}. Another possibility is to sum the contributions from individual defects located in image boxes, considering that each defect is in an infinite medium.

It is known, however, that elastic fields obtained by direct summation over periodic images can contain a spurious component if the lattice sums are not \emph{absolutely} but only \emph{conditionally} convergent~\cite{Cai2003,Kuykendall2013}. A correction scheme has been proposed by W. Cai \emph{et al.} to recover a truly periodic elastic solution~\cite{Cai2003}. Consider, for example, the strain or stress fields produced by dislocation loops and cavities in three-dimensional simulations. For both kinds of defects, elastic fields decay as $1/x^3$ for large values of $x$ due to their non-zero dipolar component, so the sum over periodic images is not absolutely convergent. Similar conditionally convergent sums are present for the calculation of the electric or magnetic fields in materials containing electric or magnetic dipoles, respectively. It is well known that for such systems, conditional convergence is related to a \emph{shape effect}~\cite{Redlack1975,Leeuw1980,Allen1987}, \emph{ie} the value of the sum depends on the shape of the summation domain. If Ewald sum, which makes use of Fourier transform, is used instead, an ``intrinsic'' value, independent of the shape, is obtained~\cite{Redlack1975}. The value given by Ewald sum can be readily deduced from a direct summation, by removing the ``extrinsic'' shape correction which only depends on the dipole moment of the simulation cell~\cite{Redlack1975,Allen1987}. However, it should be remembered that the shape effect is physical and that in general, it should be present if a finite, macroscopic system is simulated. For a material containing electric or magnetic dipoles, the extrinsic correction corresponds to the depolarization and demagnetization fields. For electrostatic problems, Ewald sum corresponds to the very particular case of a system surrounded by a medium of infinite dielectric constant (``tin foil'' boundary condition) \cite{Leeuw1980}.

For elastic problems, the situation is more complex. Contrary to electric and magnetic dipoles, the elastic field produced by an elastic dipole  is not an intrinsic property of this defect, it depends on the prescribed boundary conditions. If boundaries are sufficiently far from the defect, the elastic solution in an infinite medium is appropriate. In the vicinity of a surface, this solution, however, leads to the appearance of surface tractions which must be canceled if a free surface is considered. An additional mechanical loading can be added if necessary. The question which arises is therefore the following: which correction, if any, should be added to the elastic field computed by direct summation  to simulate the elastic field at the center of a macroscopic system of a given shape, with zero surface tractions? We will see that in general, a correction must be added, which depends on the macroscopic shape of the system. This correction also depends on the magnitude of the dipole component in the simulation box.

We start by recalling the correction proposed by W. Cai \emph{et al.} to simulate periodic systems. This correction is reformulated in as surface integral over the macroscopic system. This formulation is then used to show that applying the correction amounts to simulating a macroscopic but finite system with uniform loading, related to the elastic dipole density. A correction is proposed to simulate a macroscopic system with zero surface tractions. In the last section, the impact of elastic corrections on the kinetics of dislocation loop growth by absorption of point defects is highlighted.

\section{Reformulation of the correction for periodic systems}
\label{sec:reformulation-correction-periodic-systems}

In this section we investigate the physical meaning of the correction proposed by W. Cai \emph{et al.}~\cite{Cai2003} to remove the  component of the strain or stress field linked to non-periodicity of the displacement field, when a direct summation over images is used. We consider a three-dimensional simulation box containing a non-zero elastic dipole component. It means that far from the simulation box, the stress and strain fields produced by all defects contained in the box decay as $1/x^3$, where $x = |\bm{x}|$ and $\bm{x}$ is the position relative to center of the box. This is the case, for example, for a collection of dislocation loops and cavities. In this section we focus on the strain field. The same reasoning can be applied for the stress field. 

\subsection{Correction for periodic systems}
\label{sec:correction-periodic-systems-cai}

W. Cai \emph{et al.} have shown that absolutely convergent sums converge to a field which is periodic, so the lack of periodicity is closely linked to the conditional convergence of the direct summation on image boxes. Since absolute convergence is obtained for terms which decay as $1/x^4$, but not $1/x^3$, the first derivative of the strain field is absolutely convergent. Therefore, the strain field $\bm{\varepsilon}$ can be written, by integration of the absolutely convergent field, as
\begin{equation}
  \label{eq-strain-field-decomp-absolute-plus-correction}
  \varepsilon_{ij}({\bm{x}}) = \varepsilon_{ij}^{\mathrm{PBC}}(\bm{x}) + \varepsilon_{ij}^0,
\end{equation}
where $\varepsilon_{ij}^{\mathrm{PBC}}$ is the strain field corresponding to the periodic solution of the problem and $\varepsilon_{ij}^0$ is a contribution linked to the non-periodic character of the displacement field $\bm{u}$. By integration, this field reads
\begin{equation}
  \label{eq-displacement-field-cai}
  u_i(\bm{x}) = u_i^{\mathrm{PBC}}(\bm{x}) + \bm{g}_i\cdot \bm{x} + u_i^{0},
\end{equation}
where $u_i^{\mathrm{PBC}}$ is the periodic displacement field. It is related to $\varepsilon_{ij}^{\mathrm{PBC}}(\bm{x})$ by
\begin{equation}
  \label{eq-link-uPBC-eps-PBC}
  \varepsilon_{ij}^{\mathrm{PBC}}(\bm{x}) =  \frac{1}{2}\left(u_{i,j}^{\mathrm{PBC}}(\bm{x}) + u_{j,i}^{\mathrm{PBC}}(\bm{x}) \right),
\end{equation}
where $u_{i,j} = \partial u_i/ \partial x_j$ and $\bm{g}_i$ is a constant vector such that
\begin{equation}
  \label{eq-link-g-epsilon0}
  \varepsilon_{ij}^0= \frac{1}{2}(g_{ij}+g_{ji}).
\end{equation}

In practice, the strain field is calculated by summing over periodic images contained in a given region $\mathcal{V}$, which leads to expression~\eqref{eq-strain-field-decomp-absolute-plus-correction}. The constant field $\bm{\varepsilon}^0$ can be deduced from $\bm{g}_i$, which is computed, for example, by evaluating the displacement field at one corner of the box and at the three adjacent corners. It is important, in this case, to use the same summation domain $\mathcal{V}$. Indeed we will see in next section that  $\bm{g}_i$ and thus $\bm{\varepsilon}^0$ depend on the shape of the summation domain. These quantities, however, do not depend on the order of summation (although $u_i^0$, in general, does). To obtain the solution corresponding to a periodic system, it is necessary to subtract $\bm{\varepsilon}^0$ from $\bm{\varepsilon}$.

Such corrections are used not only in the framework of dislocation dynamics, but also in atomistic calculations to evaluate formation energies of isolated defects~\cite{Varvenne2013}. In numerical simulations, formation energies contain a spurious component due to the interaction between the defect, which can be modeled as an elastic dipole, and its periodic images. To remove this interaction energy, a direct summation of the strain field on periodic images can be performed and the component $\bm{\varepsilon}^0$ must then be subtracted. We note that in this context, other formulations for the correction of the energy have been recently derived~\cite{Pasianot2016,Dudarev2018}.

\subsection{An alternative corrective scheme using surface integrals}
\label{sec:alternative-surface-integrals}

For an infinite elastic medium, the displacement field generated by an elastic dipole $p_{jk}$, which is the first moment of a localized point-force distribution, reads~\cite{Siems1968,Clouet2018}
\begin{equation}
  \label{eq-displacement-field-elastic-dipole}
  u_i(\bm{x}) = -p_{jk}G_{ij,k}^{\infty}(\bm{\bar{x}}),
\end{equation}
where $\bm{\bar{x}} = \bm{x} - \bm{x'}$ is the vector pointing from the dipole location to the point where the field is evaluated and   $G_{ij}^{\infty}$ is the elastic Green function in an infinite body. Summation over repeated indices is implied in the following.
For an isotropic material, we have
\begin{equation}
  \label{eq-Greeen-function-isotropic-material}
  G^\infty_{ij}(\bm{\bar{x}}) = \frac{1}{8\pi(1-\nu) 2 \mu} \left( \delta_{ij} \frac{3-4\nu}{\bar{x}}+ \frac{\bar{x}_i \bar{x}_j}{\bar{x}^3}\right).
\end{equation}
In this equation, $\mu$ is the shear modulus, $\nu$ is the Poisson's ratio and  $\delta_{ij}$ is the Kronecker delta.

For a system of volume $\mathcal{V}$ containing an array of $N$ identical elastic dipoles $p_{jk}$, the displacement is
\begin{equation}
  \label{eq-displacement-field-N-dipoles}
  u_i(\bm{x}) = - \sum_{\alpha=1}^N p_{jk} G_{ij,k}^{\infty}(\bm{\bar{x}}^{(\alpha)}).
\end{equation}
This sum is evaluated inside the material, with $\bm{x} \ne \bm{x}'^{(\alpha)}$ for $\alpha = 1, \dots, N$.
For dipoles which are far from $\bm{x}$, the discrete sum can be approximated by an integral. Actually it is possible to perform this integral over the whole volume, since is is absolutely convergent for $x \rightarrow 0$ (it behaves as $1/x^2$~\cite{Leathem1912}). The result is not guaranteed to be the same as the discrete sum, but what is important is that we capture the contribution from faraway sources. The displacement is
\begin{equation}
  \label{eq-integral-displacement-N-dipoles}
  u_i(\bm{x}) = -\int_{\mathcal{V}} P_{jk} G_{ij,k}^{\infty}(\bm{\bar{x}}) \mathrm{d}V',
\end{equation}
where $P_{jk} = p_{jk}/V$ is an elastic dipole density and $V = \mathcal{V}/N$ is the volume corresponding to a single dipole. Using Ostrogradsky's theorem and the fact that $G_{ij,k}^{\infty}(\bm{\bar{x}}) = -G_{ij,k'}^{\infty}(\bm{\bar{x}})$, we obtain
\begin{equation}
  \label{eq-integral-displacement-dipoles-with-gauss}
  u_i(\bm{x}) =  \int_{\mathcal{S}} P_{jk} n_k G_{ij}^{\infty}(\bm{\bar{x}}) \mathrm{d}S',  
\end{equation}
with $\bm{n}$ the outward-pointing normal to the surface $\mathcal{S}$ which delimits $\mathcal{V}$. The elastic strain can be readily deduced:
\begin{equation}
  \label{eq-elastic-strain-from-dipole-density}
  \varepsilon_{ij}(\bm{x}) = \frac{1}{2}\int_{\mathcal{S}}  P_{lk}n_k \left[ G_{il,j}^{\infty}(\bm{\bar{x}}) + G_{jl,i}^{\infty}(\bm{\bar{x}}) \right] \mathrm{d}S'.
\end{equation}
This expression corresponds to the strain produced by surface forces $\bm{f} = \bm{P} \mathrm{d}\bm{S'}$~\cite{Leibfried1978}.

We see that the direct sum of the strain field created by dipoles in $\mathcal{V}$ has a contribution which is due to surface forces on the boundary of the summation domain. Since a periodic system has no surfaces, the contribution of these surface forces  to the strain field must be subtracted from the direct sum to recover a periodic system. It thus appears that Eq.~\eqref{eq-elastic-strain-from-dipole-density} corresponds to the spurious field $\bm{\varepsilon}^0$. The same volume must in principle be used for the discrete sum and the contribution of surface forces. Actually, since the function to integrate over $\mathcal{S}$ varies as $1/x^2$, the integral does not depend on the volume itself, but only on the \emph{shape} of the volume.

Within the framework of anisotropic elasticity, efficient numerical evaluations of the derivative of elastic Green function can be used to compute the integral~\cite{Barnett1972}. In isotropic elasticity, the integral can be written under the following form, owing to Eq.~\eqref{eq-Greeen-function-isotropic-material}:
\begin{multline}
  \label{eq-elastic-strain-from-dipole-density-isotropic}
  \varepsilon_{ij}(\bm{x}) = \frac{1}{32\pi (1-\nu) \mu} \int_{\mathcal{S}} P_{lk}n_k \left(-\delta_{il} (3-4\nu) \frac{\bar{x}_j}{\bar{x}^3} + \delta_{lj} \frac{\bar{x}_i}{\bar{x}^3} +  \delta_{ij} \frac{\bar{x}_l}{\bar{x}^3} - 3 \frac{\bar{x}_i\bar{x}_j\bar{x}_l}{\bar{x}^5}\right.\\
  \left. - \delta_{jl}(3-4\nu) \frac{\bar{x}_i}{\bar{x}^3} + \delta_{li} \frac{\bar{x}_j}{\bar{x}^3} + \delta_{ij} \frac{\bar{x}_l}{\bar{x}^3} - 3 \frac{\bar{x}_i\bar{x}_j\bar{x}_l}{\bar{x}^5} \right) \mathrm{d}S'.
\end{multline}

For a cuboid shaped box, this integral can be calculated analytically (\ref{sec:surface-integral-strain-correction}). It takes a particularly simple form when the field is estimated at $(0,0,0)$ (Eqs.~\eqref{eq-corr-eps-11-center} and~\eqref{eq-corr-eps-12-center}). This result can be used to correct a discrete sum over a cuboid-shaped domain with the simulation box at the center of the domain.

\begin{figure}[htbp]
  \centering
  \includegraphics[width=\linewidth]{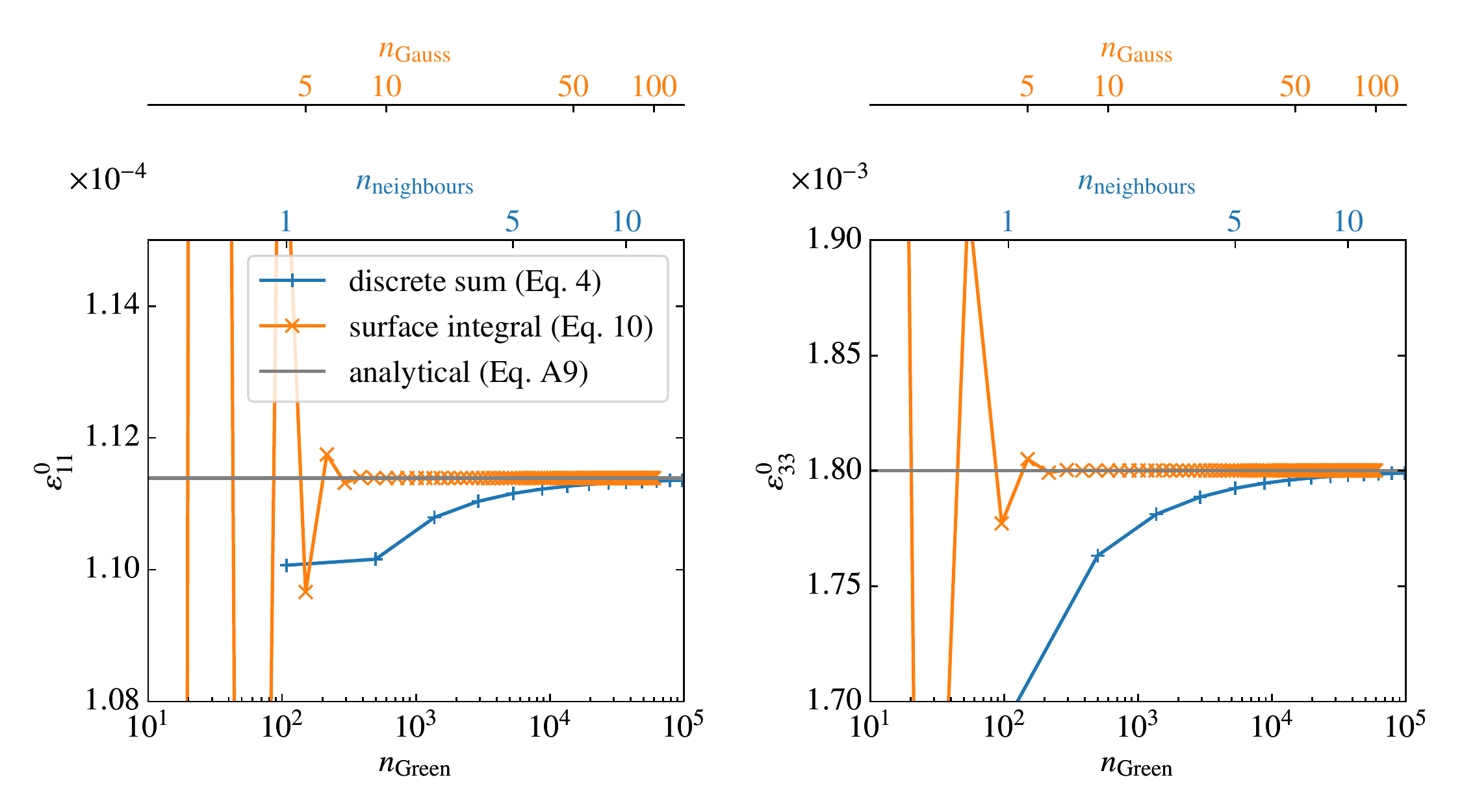}
  \caption{Comparison of the two approaches (discrete sum and surface integral) to evaluate the correction corresponding to a periodic elastic solution. The simulation box of edge length $l = 10$~nm contains an interstitial prismatic dislocation loop of radius $r=2$~nm along $x_3$ axis (see text for details). Two components are represented: (a) $\varepsilon_{11}^0$ and (b) $\varepsilon_{33}^0$. The reference solution (Eq.~\eqref{eq-corr-eps-11-center}) is shown in gray. }
  \label{fig-comp-volume-surface}
\end{figure}

To validate the explicit form of the correction given in Eq.~\eqref{eq-elastic-strain-from-dipole-density}, we consider a cubic simulation box of edge length $l = 10$~nm ($V = l^3$), containing an interstitial prismatic dislocation loop of radius $r=2$~nm along $x_3$ axis. A cuboid-shaped domain is used for the discrete sum: each image box is identified by a tuple $(n_1,n_2,n_3)$ and summation indices run from $-n_{\mathrm{neighbours}}$ to $n_{\mathrm{neighbours}}$ in the three directions. The triplet  $(0,0,0)$ corresponds to the simulation box. Isotropic elasticity is used, so that results can also be compared to the analytical solution (Eq.~\eqref{eq-corr-eps-11-center}). The elastic dipole tensor of a dislocation loop is~\cite{Clouet2018}
\begin{equation}
  \label{eq-elastic-dipole-loop}
  p_{ij} = -C_{ijkl} S_k b_l = -\mu (S_ib_j + S_jb_i) - \frac{2\nu \mu}{1-2\nu} \delta_{ij} S_k b_k,
\end{equation}
where $C_{ijkl}$ are the elastic constants, $\bm{b}$ is the Burgers vector ($\bm{b} = -b \bm{e}_{3}$) and $\bm{S}$ is the surface vector defining the area of the loop~\cite{Rovelli2018}. Here $\bm{S} = S \bm{e}_3$ with $ S = \pi r^2$. Note that $\bm{S}\cdot \bm{b} = S_k b_k = -bS$ due to the interstitial character of the loop. Parameters corresponding to aluminum are used: Burgers vector of magnitude $b = 0.2338$~nm, Poisson's ratio $\nu = 0.35$, shear modulus $\mu = 26$~GPa (not necessary for the evaluation of the strain correction). The elastic dipole density tensor is $P_{11} = P_{22} = 1.1064$~eV/nm$^3$, $P_{33} = 2.0548$~eV/nm$^3$ and $P_{ij} = 0$ for $i \neq j$, so owing to Eq.~\eqref{eq-corr-eps-12-center} components $\varepsilon_{ij}$ are also zero for $i \neq j$. In addition, $\varepsilon_{11}^0 = \varepsilon_{22}^0$, so only $\varepsilon_{11}^0$ and $\varepsilon_{33}^0$ are shown in Fig.~\ref{fig-comp-volume-surface}.

For the  approach based on direct summation of displacement fields (Eq.~\eqref{eq-link-g-epsilon0}), the displacement field is evaluated at four locations in the simulation box (the origin is at the center of the box) : $(-l/2,-l/2,-l/2)$, $(l/2,-l/2,-l/2)$, $(-l/2,l/2,-l/2)$ and $(-l/2,-l/2,l/2)$, so the number of Green function evaluations to determine the spurious strain field is $ n_{\textrm{Green}} = 4 (2n_{\mathrm{neighbours}}+1)^3$. For the surface approach (Eq.~\eqref{eq-elastic-strain-from-dipole-density}), a Gaussian quadrature is used to calculate the integral. For $n_{\textrm{Gauss}}$ integration points in one direction, the number of Green function evaluations is $n_{\textrm{Green}} = 6 n_{\mathrm{Gauss}}^2$. To compare the two methods, the strain field is represented as a function of $n_{\textrm{Green}}$. It is clear that the two approaches converge to the same result given by \eqref{eq-corr-eps-11-center} and~\eqref{eq-corr-eps-12-center}. The surface approach appears to converge faster than the direct sum approach, although in both cases values are reasonably well converged for a few hundreds of Green function evaluations, corresponding to $n_{\mathrm{neighbours}} = 1$ and $n_{\mathrm{Gauss}} = 5$. Therefore it appears that the surface method is preferable if the computation of the strain correction is required to be fast and precise. The surface method can be readily generalized to a collection of defects with non-zero elastic dipole components, once the elastic dipole density tensor is calculated.

\section{Simulation of macroscopic systems with prescribed tractions}
\label{sec:simulation-macroscopic-systems}

\begin{figure}[htbp]
  \includegraphics[width=\textwidth]{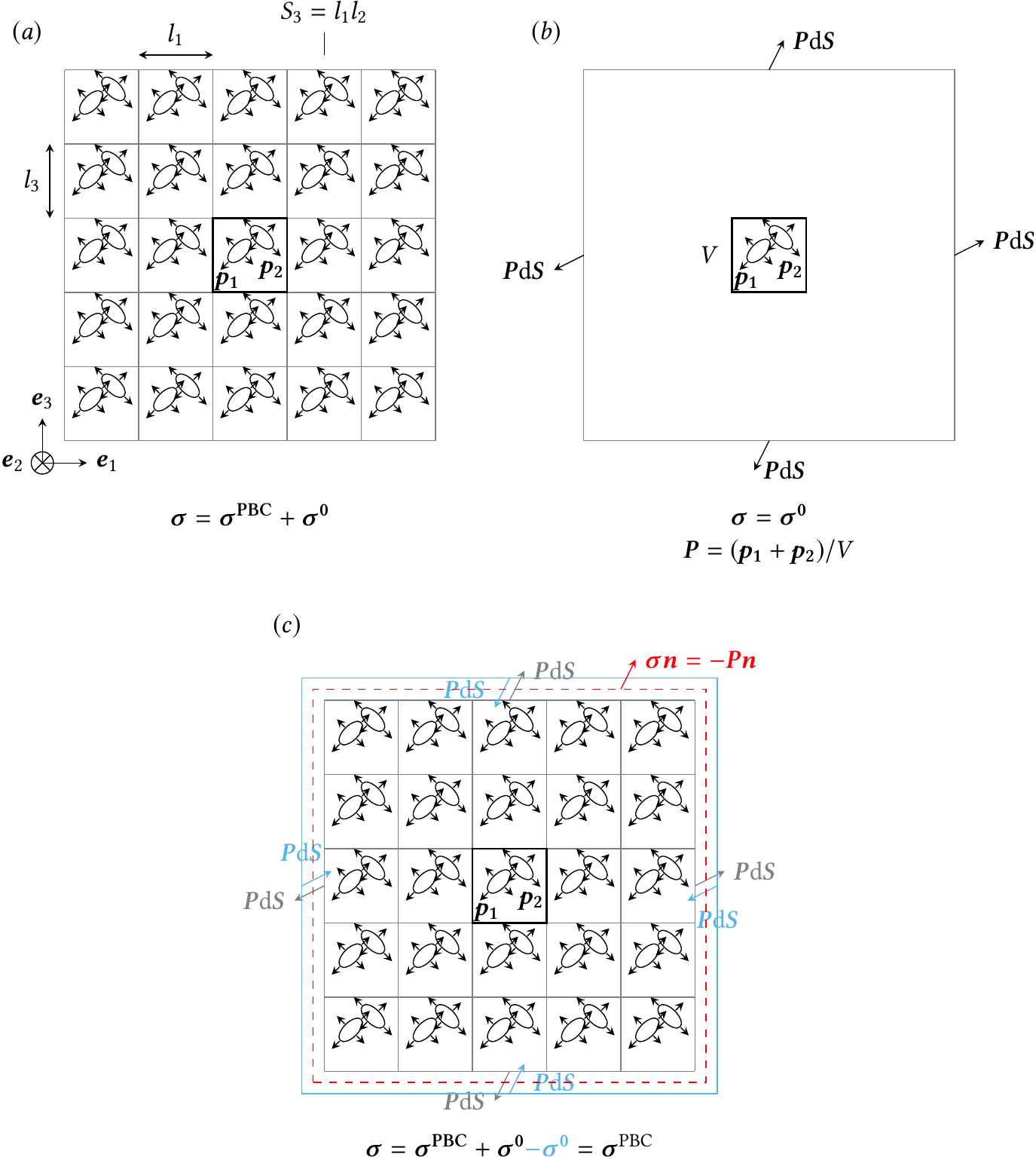}
  \caption{Schematic representation of the correction to remove the non-periodic part of the elastic solution. (a) System with periodic images: the stress field computed by direct sum over a finite set of images contains a spurious component which corresponds to a non-periodic displacement field. (b) This component can be rewritten as a contribution from surface forces $\bm{P} \mathrm{d}\bm{S}$, where $\bm{P}$ is the elastic dipole density. (c) By removing these surface forces, a periodic solution is obtained. Surface tractions on the boundary of the macroscopic system (red dashed lines) are $\bm{\sigma} \bm{n} = -\bm{P}\bm{n}$, where $\bm{n}$ is the outward-pointing normal to the surface.}
  \label{fig-schematic-compensation}
\end{figure}

We consider the case of a simulation box embedded into a macroscopic, finite system obtained by replication of the simulation box around it (Fig.~\ref{fig-schematic-compensation} (a)). Surface tractions $\bm{T} = \bm{\sigma} \bm{n}$ are imposed, where $\bm{n}$ is the normal to the surface of the system. A particular case is $\bm{T} = 0$, which corresponds to a system with free surfaces. In this section we derive the field that must be added to the discrete sum over image boxes which constitute the macroscopic system, in order to be representative of a finite system with zero surface tractions. The more general case of a given stress state can be readily obtained by adding the corresponding stress field.

To obtain the solution corresponding to zero surface tractions, a common method is to add a field which cancels surface tractions $\bm{T}$ produced by the solution for an infinite medium~\cite{Giessen1995}. This can be done, for example, by finite element (FE) solving of the elastic problem with prescribed tractions $-\bm{T}$. The traction field can be quite complicated, with steep variations on the scale of the simulation box, due to the distribution of defects in the box. However, if the simulation box is in the middle of the macroscopic system, far from surfaces, the effect of surface tractions can be accurately modelled by taking into account only their average value over a surface $S_k = l_i l_j$ defined by the box dimensions $(l_1,l_2, l_3)$. The average value over $S_k$ of the field created by a discrete set of defects of periodicity $l_i$ and $l_j$ is well approximated by integrals over a continuous distribution of dipole density, except near the edges of the system where the discrete nature of sources can be more significant.  However, such regions represent a small part of the surface and their contribution to the field in the middle of the macroscopic system is small. Higher order multipole contributions can be safely neglected if the simulation box is far from the surfaces.

It is therefore envisageable to determine the field to add to the simulation box by performing a FE solving of the elastic problem with surface tractions determined by surface integrals similar to Eq.~\eqref{eq-elastic-strain-from-dipole-density} for the stress field, evaluated outside the system, close to the surface. Actually it is possible to avoid the numerical solving phase and to derive a simple expression for this field.

As noted in the previous section, the contribution of faraway defects to the elastic field in the simulation box can be accurately described by surface forces $\bm{f}^0 = \bm{P}\mathrm{d}\bm{S'}$ (Figs.~\ref{fig-schematic-compensation} (a) and (b)). To remove the spurious field and simulate a periodic system, we have seen that we have to add to the field in the simulation box (and in the macroscopic system) the contribution from an opposite distribution of surface forces $-\bm{f}^{0} = -\bm{P}\mathrm{d}\bm{S'}$ (Fig.~\ref{fig-schematic-compensation}-(c)). The stress field, in the interspace between the two distributions of forces $\bm{f}^0$ and $-\bm{f}^0$, is such that
\begin{equation}
  \label{eq-stress-state-interspace}
  \bm{\sigma} \bm{n} = -\bm{P}\bm{n}.
\end{equation}
This expression can be obtained either by performing the integral in Eq.~\eqref{eq-elastic-strain-from-dipole-density} (in isotropic elasticity), or more simply by applying equilibrium equation of elasticity on a small volume straddling one of the two distributions of forces. The same method is used, for example, to determine the electric field due to an infinite plane of charges. The two distributions of surface forces correspond to a capacitor, where the electric field is constant if the two planes are close enough to each other.

Eq.~\eqref{eq-stress-state-interspace} means that by adding the field due to $-\bm{f}^{0} = -\bm{P}\mathrm{d}\bm{S'}$, which cancels the shape effect and leads to a periodic solution, we impose a loading of the material equal to $\bm{\sigma} = -\bm{P}$. Therefore, to cancel surface tractions, this field should be removed. This result can also be obtained directly by noting that applying the correction from Ref.~\cite{Cai2003} amounts to considering a periodic system with no imposed deformation ; it has been shown that in this case, the average stress on the simulation box, or on a group of simulation boxes, is $-\bm{P}$~\cite{Clouet2008}.

Finally, the field inside the simulation box, which corresponds to zero surface tractions on the macroscopic system, can be written as follows:
\begin{equation}
  \label{eq-field-final-with-correction-and-zero-traction}
  \bm{\sigma} = \bm{\sigma}^{\mathrm{sum}} - \bm{\sigma}^0 + \bm{P},
\end{equation}
where $\bm{\sigma}^{\mathrm{sum}} = \bm{\sigma}^{\mathrm{PBC}} + \bm{\sigma}^0$ comes from the sum over the defects in the macroscopic system, $\bm{\sigma}^0$ is the field created by surface forces $\bm{f}^0 = \bm{P}\mathrm{d}\bm{S'}$ distributed over the surface of the macroscopic system and $\bm{P}$ is the dipole density inside the simulation box. As mentioned in the previous section, $\bm{\sigma}^0$ is known analytically in isotropic elasticity and can be evaluated numerically in anisotropic elasticity by either of the two methods described in the previous section. It is important to notice that the solution with zero tractions does not depend on the shape of the sample (it is simply $\bm{\sigma} = \bm{\sigma}^{\mathrm{PBC}} + \bm{P}$). This is markedly different from the local electric and magnetic fields in systems containing electric and magnetic dipoles, which depend on the shape of the sample.

\begin{figure}[htbp]
  \includegraphics[width=\textwidth]{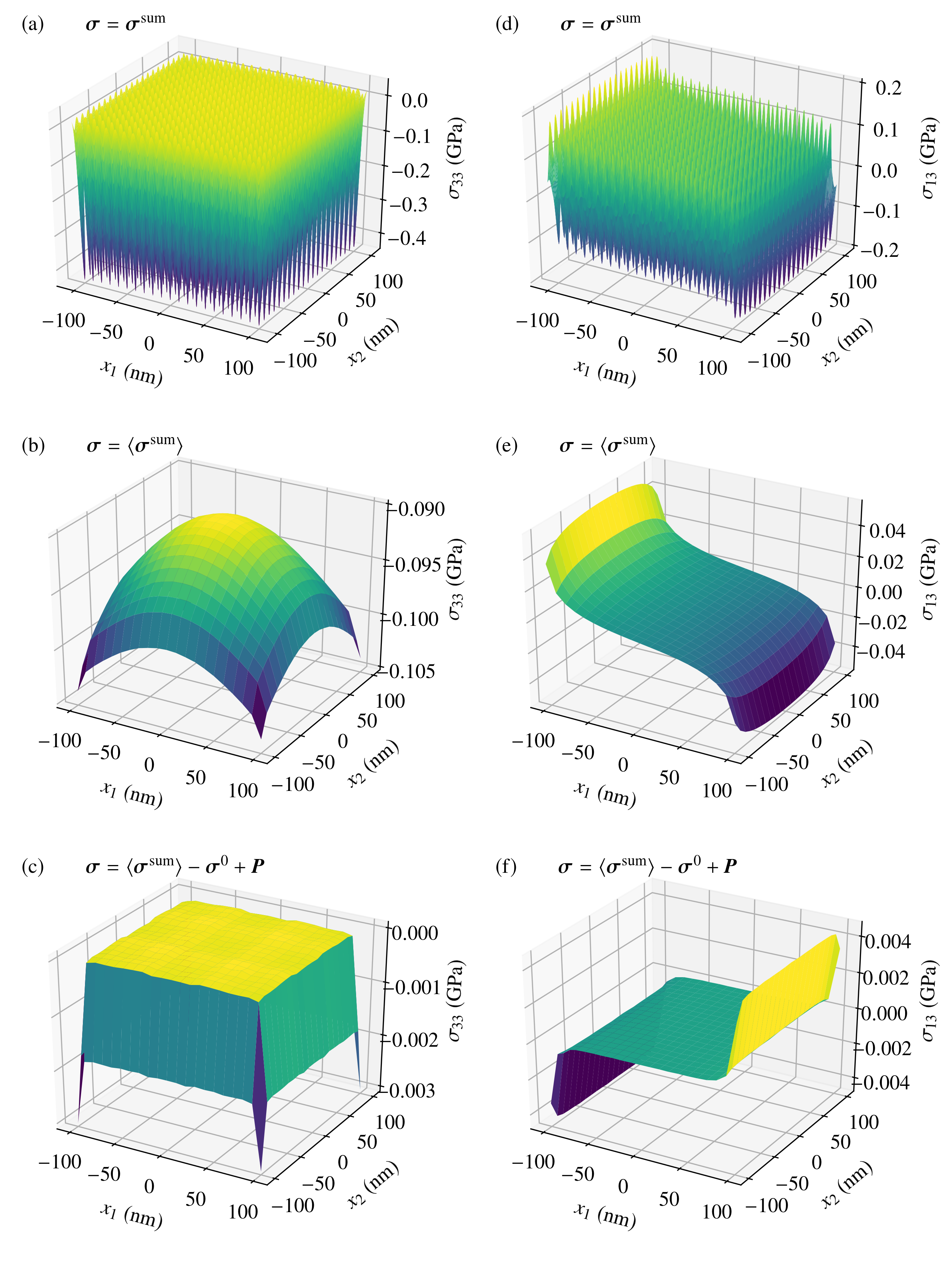}
  \caption{Stresses $\sigma_{33}$ (a-c) and $\sigma_{13}$ (d-f) on a surface of normal $[001]$ of a macroscopic system containing $21\times 21\times 21$ simulation boxes: (a,d) stress $\bm{\sigma} = \bm{\sigma}^{\mathrm{sum}}$ due to the contribution of defects inside the system (b,e) average stress $\bm{\sigma} = \langle \bm{\sigma}^{\mathrm{sum}}\rangle$ over the simulation box dimensions (c,f) average stress corrected by the non-periodic part $\bm{\sigma}^0$ and the dipole density: $\bm{\sigma} = \langle\bm{\sigma}^{\mathrm{sum}}\rangle - \bm{\sigma}^0 + \bm{P}$. The simulation box contains one prismatic loop (see text for details).}
  \label{fig-stresses-001}
\end{figure}

To validate this expression, we consider the same system as in the previous section, \emph{ie} a prismatic loop of radius $r = 2$~nm  in a cubic simulation box of edge length $l = 10$~nm. This box is duplicated 21 times along each direction to create the macroscopic system. Surface tractions produced by the solution in an infinite medium, resulting from the discrete sum over the loops, is shown in Fig.~\ref{fig-stresses-001}-(a,d) for $\sigma_{33}$ and $\sigma_{13}$ on the upper surface of normal $[001]$. They exhibit steep variations, correlated with the loop positions. However, tractions averaged over the simulation box dimensions have a much smoother profile (Fig.~\ref{fig-stresses-001}-(b,e)). This profile is mostly due to the shape effect, which can be removed by adding the field $-\bm{\sigma}^0$. By adding further the elastic dipole density $\bm{P}$, average surface tractions become essentially zero (Fig.~\ref{fig-stresses-001}-(c,f)).

\begin{figure}[htbp]
  \centering
  \includegraphics[width=0.7\textwidth]{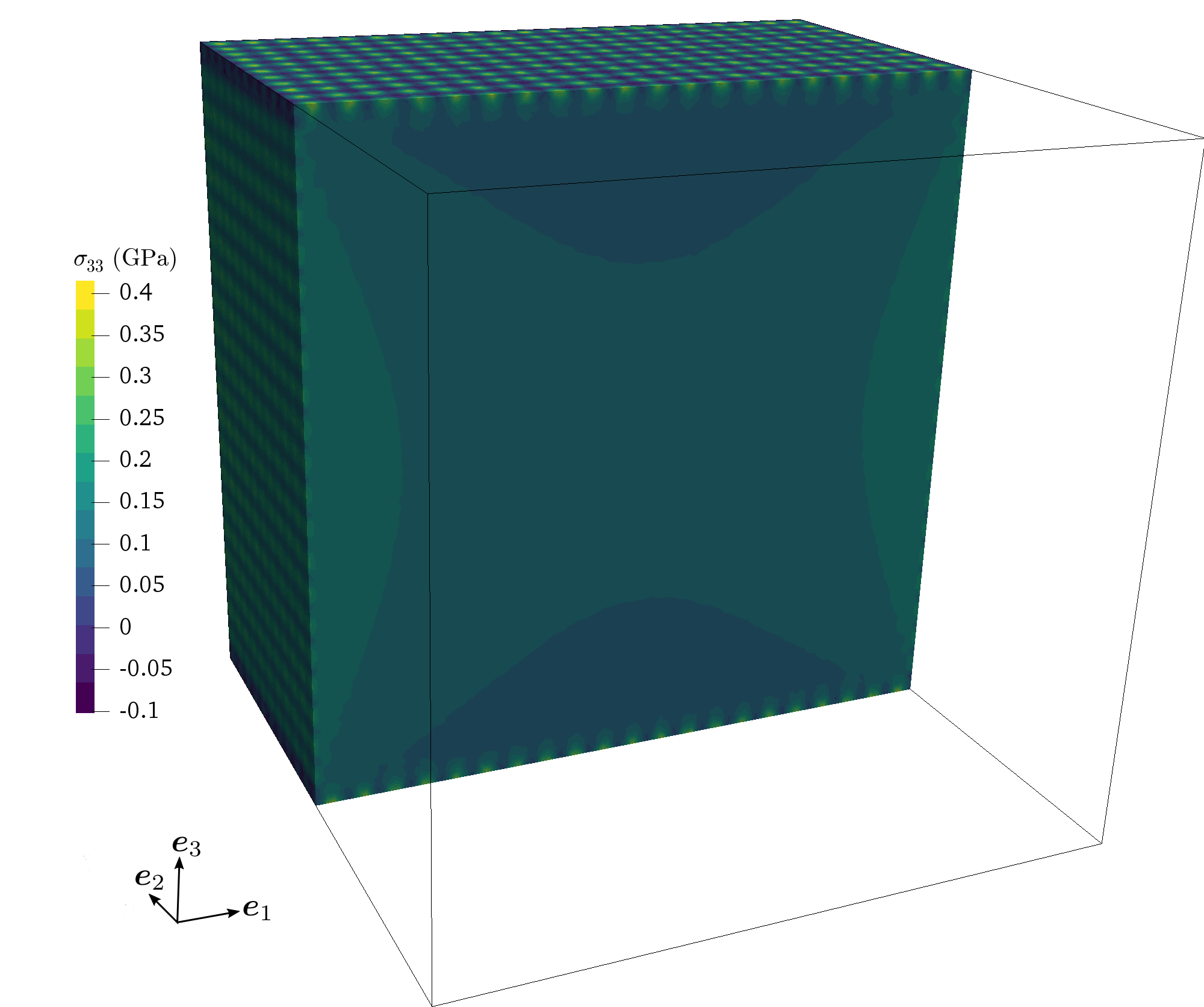}
  \caption{Stress component $\sigma_{33}$ calculated by FE modelling of a macroscopic system. Surface tractions are set to the opposite of the tractions generated by the collection of loops inside the system (see for example Fig.~\ref{fig-stresses-001}-(a,d) for one of the surfaces). The macroscopic system contains $21\times 21 \times 21$ simulation boxes. Each box has a single prismatic loop in the middle (see text for details). The system is cut half-way along $\bm{e}_{2}$ for the purpose of visualisation. }
  \label{fem_config}
\end{figure}

\begin{figure}[htbp]
  \centering
  \includegraphics[width=0.7\textwidth]{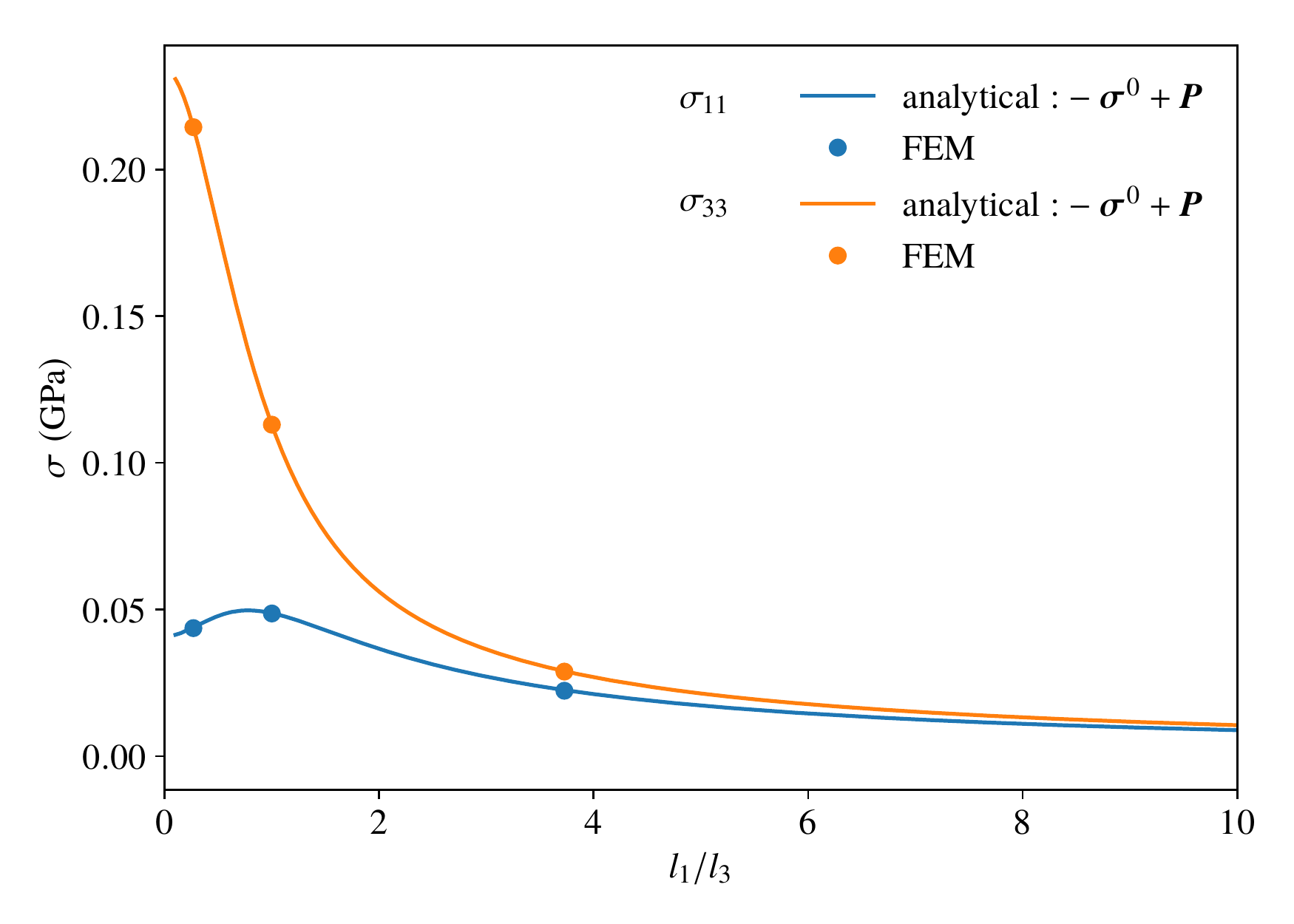}
  \caption{Correction for the stress field in the middle of a macroscopic system, corresponding to a traction-free system, for different aspect ratios $l_1/l_3$. To obtain these ratios, the macroscopic system is made of $11\times 11\times 41$, $21\times 21\times 21$ and $41\times 41\times 11$ simulation boxes. Each box has a single prismatic loop in the middle (see text for details). Reference FE simulations are compared to the analytical result derived in the present work, $\bm{\sigma} = -\bm{\sigma}^0 + \bm{P}$.}
  \label{fig-comp-fem-analytic}
\end{figure}

The accuracy of expression~\eqref{eq-field-final-with-correction-and-zero-traction} is assessed by performing reference FE calculations (see for example~\cite{Giessen1995}). Surface tractions $\bm{T}$ are obtained by summing the contributions of all the loops, as in Fig.~\ref{fig-stresses-001}-(a,d). A typical elastic solution with prescribed tractions $\bm{-T}$ is shown in Fig.~\ref{fem_config}. As noted before, in the simulation box located in the middle of the macroscopic system, the details of the surface tractions do not impact the solution, only the average value, linked to the elastic dipole density, is important. The FE solution in the middle of the macroscopic system  is compared to the analytical solution, $\bm{\sigma} = -\bm{\sigma}^0 + \bm{P}$, for different aspect ratios $l_1/l_3$ (Fig.~\ref{fig-comp-fem-analytic}). The agreement is very good, which proves that a continuous description of the traction fields, including only the dipole component, is precise enough. Corrections on $\sigma_{13}$, $\sigma_{23}$ and $\sigma_{33}$ slowly converge to zero as $l_1/l_3$ approaches infinity, since $\bm{\sigma}^0 \bm{e}_3$ approaches $\bm{P}\bm{e}_3$ in the interspace between two infinite distributions of surface forces $\pm \bm{P}\mathrm{d}S'\bm{e}_3$.

\section{Application: loop evolution under irradiation}
\label{sec:application}

In the previous section we have seen that a correction must be added to the field calculated by direct summation over near images, in order to obtain a solution corresponding to zero surface tractions on the macroscopic system. Its magnitude is proportional to the elastic dipole density, as the correction proposed in Ref.~\cite{Cai2003} which corresponds to a solution for a fully periodic system. For boxes with large elastic dipole densities, dislocation and point defect behaviours may be affected by the correction. In this section we investigate the effect of the two corrections on the loop growth under irradiation, using an object kinetic Monte Carlo (OKMC) approach~\cite{Vattre2016,Carpentier2017}.

As before, we use typical parameters for aluminum (see section~\ref{sec:alternative-surface-integrals}). Two interstitial Frank loops of different radii (2 and 3 nm) are introduced in a cubic box of edge length $l = 10$~nm, at $(l/2,l/2,l/4)$ and $(l/2,l/2,3l/4)$. The normal to their habit plane is $\bm{e}_3$. Six vacancies and self-interstitials are introduced in the box per second, which corresponds to damage rate of $10^{-4}$~dpa/s (displacements per atom). They diffuse in the simulation box with periodic boundary conditions until they are absorbed by one of the loops. Simulations are performed at $T = 300$~K. The emission of point defects by loops can be neglected at this temperature.

The migration of point defects occurs by successive hops between stable positions in the lattice. The jump frequency is given by
\begin{equation}
  \label{eq-jump-frequency}
  \nu = \nu_0 \exp{\left(-\frac{E^{\mathrm{m}}}{k_{\mathrm{B}}T}\right)},
\end{equation}
where $\nu_0= 10^{13}$~Hz is an attempt frequency and $E^{\mathrm{m}}$ is the migration energy for the considered jump. It reads~\cite{Siems1968}
\begin{equation}
  \label{eq-migration-energy}
  E^{\mathrm{m}} = E^{\mathrm{m}}_0 - p_{ij}^{\mathrm{sad}}\varepsilon_{ij} + p_{ij}^{\mathrm{sta}}\varepsilon_{ij},
\end{equation}
with $E^{\mathrm{m}}_0$ the migration energy without any strain, $p_{ij}^{\mathrm{sta}}$ and $p_{ij}^{\mathrm{sad}}$ the elastic dipoles at the stable and saddle positions and $\varepsilon_{ij}$ the strain field, which is assumed to be the same at both positions. We also suppose here that the elastic dipoles do not depend on the strain, \emph{ie} polarizability effects are not considered~\cite{Schober1984}. Elastic dipoles of vacancies and self-interstitials in aluminum at stable and saddle configurations can be found in Ref.~\cite{Carpentier2017}. 

If the local strain field is corrected by $\bm{\varepsilon}^{\mathrm{corr}}$, the migration barrier becomes
\begin{equation}
  \label{eq-migration-energy-corr}
  E^{\mathrm{m}} = E^{\mathrm{m}}_0 - (p_{ij}^{\mathrm{sad}} - p_{ij}^{\mathrm{sta}})\varepsilon_{ij} - (p_{ij}^{\mathrm{sad}} - p_{ij}^{\mathrm{sta}})\varepsilon_{ij}^{\mathrm{corr}}.
\end{equation}
Since in general elastic dipoles are not equal at stable and saddle positions, the elastic correction can alter the point defect diffusion. In particular, we can expect an effect of the correction in the simulation of phenomena such as void swelling or irradiation creep, for which the influence of the elastic field created by dislocations and cavities on the diffusion of point defects is important~\cite{Heald1975}. If the magnitude of $\varepsilon_{ij}^{\mathrm{corr}}$ is the same as $\varepsilon_{ij}$, results could change appreciably.

\begin{figure}[htbp]
  \centering
  \includegraphics[width=0.7\textwidth]{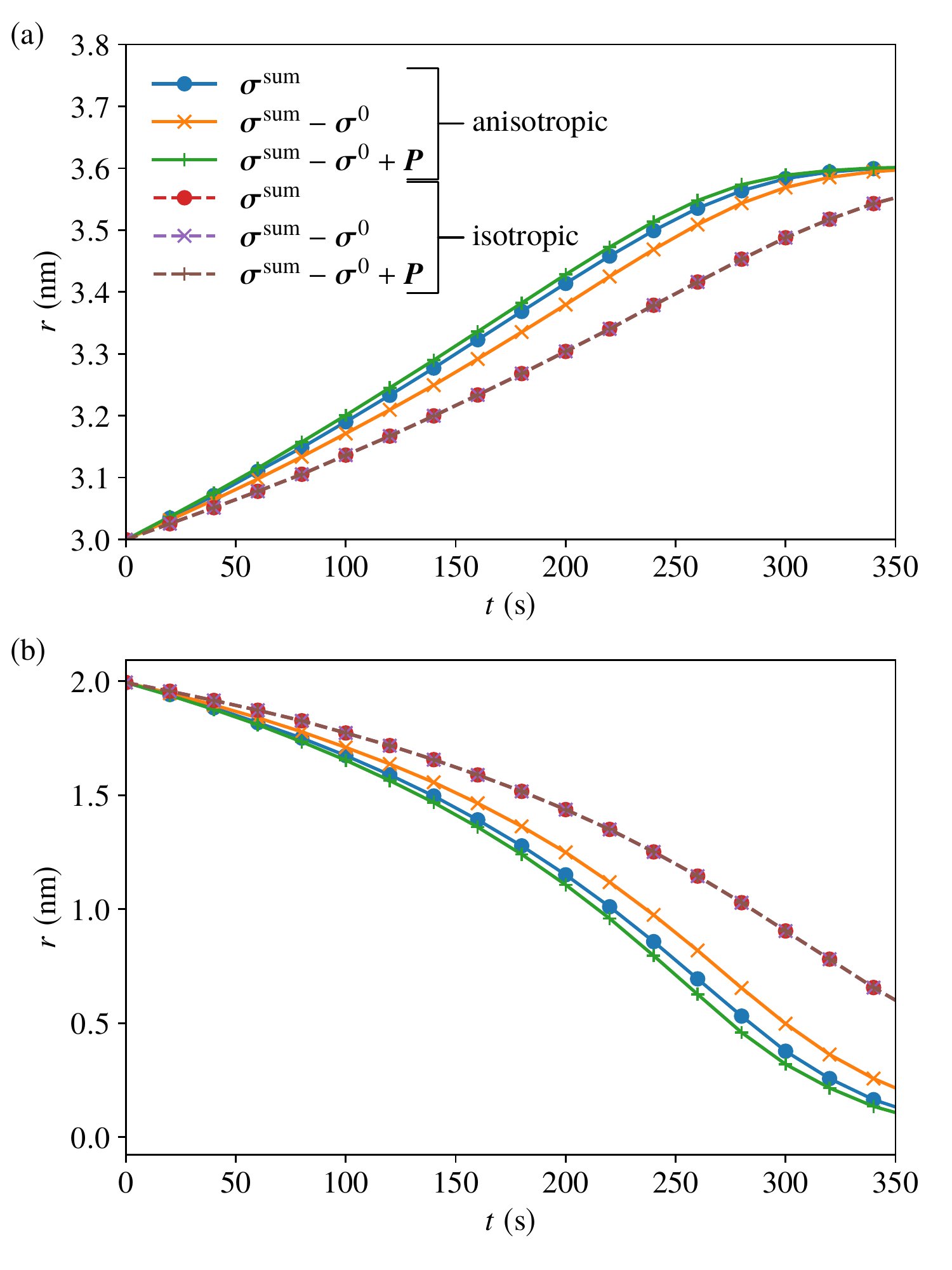}
  \caption{Evolution of the radius of two Frank loops of initial radii (a) 3 nm and (b) 2 nm in a cubic box of edge length $l = 10$~nm. Vacancies and self-interstitials are introduced simultaneously, simulating an electron irradiation with a damage rate equal to $10^{-4}$~dpa/s. Corrections $-\bm{\sigma}^0$ and $-\bm{\sigma}^0 + \bm{P}$ are added to the stress field calculated as a sum of contributions from nearby image boxes ($\bm{\sigma}^{\mathrm{sum}}$), to account for different boundary conditions. ``Anisotropic'' case corresponds to elastic dipoles obtained from DFT calculations. For the ``isotropic'' case, dipoles are assumed to be the same at stable and saddle points and are purely hydrostatic. Their trace is given by DFT results at stable position.}
  \label{fig-loop-evolution}
\end{figure}

The evolution of the two loops is given in Fig.~\ref{fig-loop-evolution} with different elastic corrections, averaged over 1000 simulations for each condition. Whatever the correction, one sees that the larger loop grows, while the smaller loop shrinks. This result is in agreement with bias calculations on single loops, which show that the bias increases with loop size~\cite{Bullough1981,Jourdan2015,Rouchette2015}. Differences in loop evolution are clearly observed for the various corrections envisaged, although they remain small from an experimental point of view. Loops exhibit the fastest evolution if the correction for zero surface tractions is used. The slowest evolution is obtained with the correction for a fully periodic system. Simulations were also performed with simplified elastic dipoles. They were taken purely hydrostatic, with the same value at stable and saddle positions, deduced from the trace of the DFT dipole tensors at stable position. No difference in loop evolution is seen in this case, in agreement with Eq.~\eqref{eq-migration-energy-corr}.

Since vacancies and interstitials are produced at the same rate, the number of interstitials in loops stays constant, providing no defects remain in the matrix. This means that the elastic correction is also roughly the same at any time. The stress correction for a fully periodic simulation is $\sigma_{11} = \sigma_{22} = 0.290$~GPa and $\sigma_{33} = 0.489$~GPa , while for a macroscopic system with zero tractions it is $\sigma_{11} = \sigma_{22} = -0.159$~GPa and $\sigma_{33} = -0.368$~GPa. These stress levels are quite substantial and could also affect processes such as dislocation glide. It therefore appears crucial to apply the stress correction corresponding to the desired boundary conditions.

\section{Conclusions}
\label{sec:conclusions}

From a simulation box with a non-zero elastic dipole component, the aim of this work is to determine the effect, in the box, of prescribed tractions at the boundary of a macroscopic system built by replicating the simulation box around it. The starting point is a reformulation of the correction proposed by Cai \emph{et al.}~\cite{Cai2003} to obtain a fully periodic elastic solution. Using this formulation, based on surface integrals, we show that the correction only depends on the shape of the macroscopic system, and that applying the correction is equivalent to simulating a macroscopic but finite system with surface tractions $-\bm{P}\bm{n}$, where $\bm{P}$ is the elastic dipole density and $\bm{n}$ an outward-pointing normal unit vector. By removing these tractions, a system containing a homogeneous distribution of defects, with zero surface tractions, is simulated. The elastic solution thus obtained does not depend anymore on the shape of the macroscopic system.

Elastic corrections are applied in OKMC simulation boxes to simulate the evolution of dislocation loops under irradiation, due to the absorption of point defects. It is shown that the dislocation loop evolution depends on the correction if point defects have different properties at stable and saddle points. It can be expected that these elastic corrections not only have an influence on point defect diffusion, but also on dislocation movement. Therefore it appears important to be aware of the type of system that is simulated when elastic corrections are applied, and to apply the desired correction.

\section{Acknowledgments}
\label{sec:acknowledgments}

The author is grateful to E. Clouet for useful discussions and for his comments on the manuscript. Part of this work has been carried out within the framework of the EUROfusion Consortium and has received funding from the Euratom research and training programme 2014-2018 under Grant Agreements No. 633053.  The views and opinions expressed herein do not necessarily reflect those of the European Commission.

\appendix

\section{Surface integrals for the strain correction}
\label{sec:surface-integral-strain-correction}

The analytical form of the strain field which must be subtracted from a direct sum to obtain a fully periodic solution is given here in the case of isotropic elasticity, for a cuboid shaped box of dimensions $(l_1,l_2,l_3)$. The origin is at the center of the box. Expressions are only given for $\varepsilon_{11}^0$ and $\varepsilon_{12}^0$, other terms are obtained by cyclic permutation of indices. 

\begin{multline}
  \label{eq-corr-eps-11-gen}
  \varepsilon_{11}^0(\bm{x}) = \frac{1}{16 \pi (1-\nu) \mu}\sum_{\substack{u = 0,1 \\ v = 0,1 \\ w = 0, 1}}  (-1)^{u+v+w} \left\{ \phantom{\frac{l}{2}}  \right.\\
    \left.  P_{11} \left[-2(1-2\nu) A\left(x_1-(-1)^u \frac{l_1}{2},x_2-(-1)^v \frac{l_2}{2},x_3-(-1)^w \frac{l_3}{2}\right) \right.\right.\\
  \left. - B\left(x_1-(-1)^u \frac{l_1}{2},x_2-(-1)^v \frac{l_2}{2},x_3-(-1)^w \frac{l_3}{2}\right)\right]  \\
  + P_{22} E\left(x_2-(-1)^v \frac{l_2}{2}, x_1-(-1)^u \frac{l_1}{2},x_3-(-1)^w \frac{l_3}{2}\right) \\
  + P_{33} E\left(x_3-(-1)^w \frac{l_3}{2}, x_1-(-1)^u \frac{l_1}{2}, x_2-(-1)^v \frac{l_2}{2}\right) \\
  - 2P_{23} F\left(x_2-(-1)^v \frac{l_2}{2}, x_3-(-1)^w \frac{l_3}{2}, x_1-(-1)^u \frac{l_1}{2}\right) \\
  - P_{13} \left[2(1-2\nu) D\left(x_1-(-1)^u \frac{l_1}{2},x_3-(-1)^w \frac{l_3}{2},x_2-(-1)^v \frac{l_2}{2} \right) \right. \\
  \left. + 2 C\left(x_1-(-1)^u \frac{l_1}{2},x_3-(-1)^w \frac{l_3}{2},x_2-(-1)^v \frac{l_2}{2} \right) \right] \\
  - P_{12} \left[2(1-2\nu) D\left(x_1-(-1)^u \frac{l_1}{2},x_2-(-1)^v \frac{l_2}{2},x_3-(-1)^w \frac{l_3}{2} \right) \right. \\
  \left.\left. + 2 C\left(x_1-(-1)^u \frac{l_1}{2},x_2-(-1)^v \frac{l_2}{2},x_3-(-1)^w \frac{l_3}{2} \right) \right]\right\}
\end{multline}

\begin{multline}
  \label{eq-corr-eps-12-gen}
  \varepsilon_{12}^0(\bm{x}) = \frac{1}{16 \pi (1-\nu) \mu}\sum_{\substack{u = 0,1 \\ v = 0,1 \\ w = 0, 1}}  (-1)^{u+v+w} \left\{ \phantom{\frac{l}{2}} \right. \\
  \left. P_{11} \left[-(1-2\nu) D\left(x_1-(-1)^u \frac{l_1}{2},x_2-(-1)^v \frac{l_2}{2},x_3-(-1)^w \frac{l_3}{2}\right) \right. \right. \\
  \left. +  C\left(x_1-(-1)^u \frac{l_1}{2},x_2-(-1)^v \frac{l_2}{2},x_3-(-1)^w \frac{l_3}{2}\right)\right] \\
  + P_{22} \left[-(1-2\nu) D\left(x_2-(-1)^v \frac{l_2}{2},x_1-(-1)^u \frac{l_1}{2},x_3-(-1)^w \frac{l_3}{2}\right) \right. \\
  \left. +  C\left(x_2-(-1)^v \frac{l_2}{2},x_1-(-1)^u \frac{l_1}{2},x_3-(-1)^w \frac{l_3}{2}\right)\right] \\
  - P_{33} F\left(x_1-(-1)^u \frac{l_1}{2},x_2-(-1)^v \frac{l_2}{2},x_3-(-1)^w \frac{l_3}{2}\right) \\
  + P_{23} \left[-2(1-\nu) D\left(x_3-(-1)^w \frac{l_3}{2},x_1-(-1)^u \frac{l_1}{2},x_2-(-1)^v \frac{l_2}{2}\right) \right. \\
  \left. -2 F\left(x_3-(-1)^w \frac{l_3}{2},x_1-(-1)^u \frac{l_1}{2},x_2-(-1)^v \frac{l_2}{2}\right) \right] \\
  + P_{13} \left[-2(1-\nu) D\left(x_3-(-1)^w \frac{l_3}{2},x_2-(-1)^v \frac{l_2}{2},x_1-(-1)^u \frac{l_1}{2}\right) \right. \\
  \left. -2 F\left(x_3-(-1)^w \frac{l_3}{2},x_2-(-1)^v \frac{l_2}{2},x_1-(-1)^u \frac{l_1}{2}\right) \right] \\
  + P_{12} \left[-2(1-\nu) A\left(x_2-(-1)^v \frac{l_2}{2},x_1-(-1)^u \frac{l_1}{2},x_3-(-1)^w \frac{l_3}{2}\right) \right. \\
  +  E\left(x_2-(-1)^v \frac{l_2}{2},x_1-(-1)^u \frac{l_1}{2},x_3-(-1)^w \frac{l_3}{2}\right) \\
  - 2(1-\nu) A\left(x_1-(-1)^u \frac{l_1}{2},x_2-(-1)^v \frac{l_2}{2},x_3-(-1)^w \frac{l_3}{2}\right) \\
  \left. \left. +  E\left(x_1-(-1)^u \frac{l_1}{2},x_2-(-1)^v \frac{l_2}{2},x_3-(-1)^w \frac{l_3}{2}\right) \right] \right\}
\end{multline}

Functions used in Eqs~\eqref{eq-corr-eps-11-gen} and~\eqref{eq-corr-eps-12-gen} are given by
\begin{align}
  \label{eq-A-func}
  A(x,y,z) &= \arctan{\left(\frac{yz}{x\sqrt{x^2+y^2+z^2}}\right)} \\
  \label{eq-B-func}
  B(x,y,z) &= \frac{xyz(2x^2+y^2+z^2)}{(x^2+y^2)(x^2+z^2)\sqrt{x^2+y^2+z^2}} \\
  \label{eq-C-func}
  C(x,y,z) &= \frac{x^2z}{(x^2+y^2)\sqrt{x^2+y^2+z^2}} \\
  \label{eq-D-func}
  D(x,y,z) &= -\ln{\left(z+\sqrt{x^2+y^2+z^2} \right)} \\
  \label{eq-E-func}
  E(x,y,z) &= \frac{xyz}{(x^2+y^2)\sqrt{x^2+y^2+z^2}} \\
  \label{eq-F-func}
  F(x,y,z) &= \frac{z}{\sqrt{x^2+y^2+z^2}}
\end{align}

When a simulation box is embedded into a macroscopic system, it is natural to place the simulation box in the middle of the system. The field must then be evaluated at the center of the system (\emph{ie} $\bm{x} = (0,0,0)$). Only functions which are odd in $x$, $y$ and $z$ contribute to the result, so the field takes the remarkably simple form:
\begin{multline}
  \label{eq-corr-eps-11-center}
  \varepsilon_{11}^0(0) =  \frac{P_{11}}{2\pi(1-\nu) \mu} \left[2(1-2\nu) A(l_1,l_2,l_3) + B(l_1,l_2,l_3)\right] \\
  - \frac{P_{22}}{2\pi (1-\nu) \mu} E(l_2,l_1,l_3) - \frac{P_{33}}{2\pi (1-\nu) \mu} E(l_3,l_1,l_2)
\end{multline}
\begin{multline}
  \label{eq-corr-eps-12-center}
  \varepsilon_{12}^0(0) = \frac{P_{12}}{2\pi(1-\nu)\mu} \left[ 2(1-\nu) \left(A(l_1,l_2,l_3) + A(l_2,l_1,l_3) \right)- \left(  E(l_1,l_2,l_3) + E(l_2,l_1,l_3)\right)\right]
\end{multline}

\bibliographystyle{model1-num-names}
\bibliography{biblio}

\end{document}